\documentclass[12pt]{article}
\usepackage{amssymb,amsmath}
\newcommand{\nin}{\noindent}
\newcommand{\be}{\begin{equation}}
\newcommand{\ee}{\end{equation}}
\newcommand{\bea}{\begin{eqnarray}}
\newcommand{\eea}{\end{eqnarray}}
\newcommand{\br}{\hskip .25cm/\hskip -.25cm}
\newcommand{\nn}{\nonumber\\}

\newcommand{\ol}{\overline}
\usepackage{graphicx}  
\begin{document}

\begin{center}

{\Large{\bf Control of quantum fluctuations for a Yukawa interaction in the Kaluza Klein picture}}

\vspace{1cm}

{\bf J. Alexandre}$^{1}$ and {\bf K. Farakos}$^{2}$

\vspace{0.5cm}

{\it
1 Department of Physics, King's College London, WC2R 2LS, UK.

2 Department of Physics, National Technical University of Athens,\\
Zografou Campus, 15780 Athens, Greece}

\vspace{1cm}

{\bf Abstract}

\end{center}

We study a system of fermions interacting with a scalar field, in 4+1 dimensions where the 5th
dimension is compactified, using an exact functional method, where quantum fluctuations are 
controlled by the amplitude of the bare fermion mass. The integration of our equations leads to the
properties of the dressed Yukawa coupling, that we study at one-loop so as to show the 
consistency of the approach.
Beyond one loop, the non-perturbative aspect of the method gives us the possibility to 
derive the dynamical fermion mass. The result obtained is cut off independent
and this derivation proposes an alternative to the Schwinger-Dyson approach.

\vspace{1cm}

\section{Introduction}

Motivated by the unification of fundamental interactions, the Kaluza Klein framework has long been 
considered, where additional dimensions are compactified, on a TeV scale. These energy scales are 
inaccessible to present experiments, but the next generation of particle accelerators could give an insight
into these high energy processes.
We propose in this context a functional method in order to built the quantum theory corresponding to
a Yukawa interaction, in 4+1 dimensions, where the 5th dimension is compactified.
The idea is to control quantum fluctuations with the bare fermion mass of the theory,
whose amplitude is controlled by a dimensionless parameter $\lambda$.
If $\lambda\to\infty$, the bare system deals with infinitely massive fermions which therefore decouple from the 
dynamics: the system describes a self interacting scalar field. 
As $\lambda$ decreases, quantum fluctuations arising from the fermions gradually appear and the full quantum
system sets up. The interesting point is that it is possible to derive an exact evolution
equation for the proper graphs generator functional with $\lambda$. The corresponding flows
can be seen as renormalization flows, different from Wilsonian flows, though: the bare mass
term, quadratic in the fermion field, does not play the role of a regulator and we need, in addition,
a cut off so as to define the $\lambda$-flows.  

Despite the difference with Wilsonian flows, the $\lambda$-flows recovers the usual one-loop
renormalization results. Beyond one loop, these flows are
derived within a given gradient expansion which is assumed for the functional dependence of the
quantum theory, instead of a loop expansion. The gradient expansion provides a ressumation
of a sub-class of graphs and is non-perturbative.
This method was successfully applied to a scalar model \cite{scalar} and in $QED$ \cite{QED}, 
in planar systems \cite{moi} and for the Coulomb interaction in an electron gas \cite{gas}. 
Reviews can be found in \cite{reviewjanos,8essalovikn}.

We present the model in section 2, and we derive the different evolution equations.
We discuss in section 3 the possibility of having a massless scalar field, for which the restoration force is 
provided by the scalar self interaction. In this case, the bare scalar mass has to be chosen carefully and we show that
this implies an inequality between the Yukawa coupling and the scalar self coupling, in order not to
generate a spontaneous symmetry breaking.
Section 4 deals with the properties of the four dimensional Yukawa coupling at one loop, 
and show the consistency of the approach. The Yukawa coupling is obtained as an explicit function of three
mass scales: the bare fermion mass, the compactification radius and the cut off. These mass scales 
are independent, which enables us to take several limits. For example, 
for fixed cut off and fixed compactification radius, the quantum theory recovers the classical one
in the limit where $\lambda\to\infty$, as expected. We consider then in section 5
the dynamical generation of a fermion mass, as a result of the non perturbative aspect of our evolution 
equations. This study is possible in the functional method presented here by allowing the parameter 
$\lambda$ to go to 0, for fixed cut off and compactification radius. 
The result is consistent with existing literature, and has the advantage
to be independent of any renormalization scheme, since the cut off dependence disappears 
when $\lambda\to 0$. Finally, section 6 contains our conclusions and future work.

\section{Model and evolution equations}

For the sake of clarity and in order to explain the approach used here, we describe
a system made of fermions coupled to a real scalar field. I order to describe a parity-invariant theory,
we consider two four-component fermion flavours, that are written as one eight-component fermion 
$\psi=(\psi^{(1)},\psi^{(2)})$. We have then
$8\times 8$ gamma matrices which are
diagonal in the flavour space and read $\Gamma^\mu=diag(\gamma^\mu,-\gamma^\mu)$, $\mu=0,...,4$, 
where the matrices $\gamma^\rho$, $\rho=0,...,3$ are the irreducible four dimensional gamma matrices,
and $\gamma^4=\gamma^0\gamma^1\gamma^2\gamma^3$ \cite{gamma}.  

The bare action is:
\be\label{bareaction}
S=\int d^5x\left\{\frac{1}{2}\partial_\mu\phi\partial^\mu\phi+i\ol\psi\br\partial\psi-g_0\phi\ol\psi\psi
-\lambda m_0\ol\psi\psi-U^B_0(\phi)\right\},
\ee
where $U_0^B(\phi)$ is a bare scalar potential, and the partition function is:
\be
Z=\int{\cal D}[\Psi,\Phi]\exp\left(iS+i\int d^5x(\ol\Psi\eta+\ol\eta\Psi+j\Phi)\right),
\ee
where $\ol\eta,\eta,j$ are the sources for $\Psi,\ol\Psi,\Phi$ respectively.
The derivatives of the connected graphs generator functional $W=-i\ln Z$ define the classical fields:
\bea\label{derivativesW}
\frac{\delta W}{\delta\ol\eta}&=&\left<\Psi\right>=\psi\\
\frac{\delta W}{\delta\eta}&=&-\left<\ol\Psi\right>=-\ol\psi\nn
\frac{\delta W}{\delta j}&=&\left<\Phi\right>=\phi\nn
\frac{\delta^2 W}{\delta\ol\eta\delta\eta}&=&i\ol\psi\psi-i\left<\ol\Psi\Psi\right>\nonumber,
\eea
where
\be
\left<\cdot\cdot\cdot\right>=\frac{1}{Z}\int{\cal D}[\Psi,\Phi](\cdot\cdot\cdot)
\exp\left(iS+i\int d^5x(\ol\Psi\eta+\ol\eta\Psi+j\Phi)\right).
\ee
The proper graphs generator functional $\Gamma$ is defined as the Legendre transform of $W$:
\be
\Gamma=W-\int d^5x\left(\ol\psi\eta+\ol\eta\psi+j\phi\right),
\ee
and has $\lambda,\ol\psi,\psi,\phi$ as independent variables. 
$\Gamma$ depends on $\lambda$, which therefore
parametrizes a family of quantum theories.
When $\lambda\to\infty$, the fermions become infinitely massive and decouple from the
dynamics: the effective action describes a self interacting scalar field.
When $\lambda$ decreases, quantum fluctuations gradually appear in the fermion dynamics and
the couplings fermion/scalar get dressed. Finally,
the full quantum effects are present in $\Gamma$ when $\lambda$ decreases down to 1 (for a massive
bare theory) or 0 (for a massless bare theory).

\subsection{Evolution equation for $\Gamma$}

The functional derivatives of $\Gamma$ are:
\bea\label{derivativesG}
\frac{\delta\Gamma}{\delta\ol\psi}&=&-\eta\\
\frac{\delta\Gamma}{\delta\psi}&=&\ol\eta\nn
\frac{\delta\Gamma}{\delta\phi}&=&-j\nn
\frac{\delta^2\Gamma}{\delta\ol\psi\delta\psi}&=&
-\left(\delta^2 W\right)^{-1}_{\ol\eta\eta},\nonumber
\eea
and we also have (remember that the independent variables of $\Gamma$ are $\lambda,\ol\psi,\psi,\phi$) 
\bea
\dot\Gamma&=&\dot W+\int d^5x\left\{\dot{\ol\eta}\frac{\delta W}{\delta\ol\eta}
-\frac{\delta W}{\delta\eta}\dot\eta+\frac{\delta W}{\delta j}\dot{j}\right\}\nn
&&~~~~-\int d^5x\left\{\ol\psi\dot\eta+\dot{\ol\eta}\psi+\dot{j}\phi\right\}\nn 
&=&\dot W,
\eea
where a dot denotes a derivative with respect to $\lambda$.
From the relations (\ref{derivativesW}), the evolution equation for $W$ with $\lambda$ is
\bea
\dot W&=&-m_0\int d^5x\left<\ol\Psi\Psi\right>\\
&=&-m_0\int d^5x\left(\ol\psi\psi+i\frac{\delta^2 W}{\delta\ol\eta\delta\eta}\right).\nonumber
\eea
From eqs.(\ref{derivativesG}), the evolution equation for $\Gamma$ is finally
\be\label{evolgamma}
\dot\Gamma+m_0\int d^5x~\ol\psi\psi=im_0\mbox{Tr}\left\{\left(\delta^2\Gamma\right)^{-1}_{\ol\psi\psi}\right\},
\ee
where $\delta^2\Gamma$ is the second derivative matrix
\be
\delta^2\Gamma=\begin{pmatrix}
\delta^2\Gamma_{\ol\psi\psi}&\delta^2\Gamma_{\ol\psi\ol\psi}&
\delta^2\Gamma_{\ol\psi\phi}\\
\delta^2\Gamma_{\psi\psi}&\delta^2\Gamma_{\psi\ol\psi}&
\delta^2\Gamma_{\psi\phi}\\
\delta^2\Gamma_{\phi\psi}&\delta^2\Gamma_{\phi\ol\psi}&
\delta^2\Gamma_{\phi\phi}
\end{pmatrix},
\ee
and $\left(\delta^2\Gamma\right)^{-1}_{\ol\psi\psi}$ is the $\ol\psi\psi$ element of its inverse.
The evolution equation (\ref{evolgamma}) is non-perturbative and provides a ressumation to all orders in $\hbar$.
The trace which appears in eq.(\ref{evolgamma}) (continuous summation over the
Fourier modes $p_0,p_1,p_2,p_3$ and discreet summation over the Kaluza-Klein modes $p_4$)
contains the quantum corrections to the classical theory and needs to be regularized
(we will consider here a cut off regulator).
This trace is similar to the one that we would obtain in a one-loop computation, but in the
present case, the action appearing in this trace is the effective action, i.e. it contains the
{\it dressed} couplings, and not the bare ones as one would have in a one-loop computation.
This feature is at the origin
of the ressumation provided by the evolution equation (\ref{evolgamma}), and is similar to
exact renormalization procedures.
                                                                                                                             
So as to take physical informations out of the evolution equation (\ref{evolgamma}), we shall
assume the following functional dependence for $\Gamma$:
\be\label{ansatz}
\Gamma=\int d^5x\left\{\frac{1}{2}\partial_\mu\phi\partial^\mu\phi+i\ol\psi\br\partial\psi
-\sum_{n\ge 0} U_n(\phi)(\ol\psi\psi)^n\right\},
\ee
where the potentials $U_n(\phi)$ depend on $\lambda$.
This ansatz corresponds to a local potential approximation
in the framework of the gradient expansion, where:
\begin{itemize}
\item $U_0(\phi)$ is the scalar potential;
\item $U_1(0)$ is the fermion mass;
\item $\partial_\phi U_1(0)$ is the Yukawa coupling;
\item $U_2(0)$ is the four-fermion interaction; etc...
\end{itemize}
and the latter quantities depend on $\lambda$,
which controls the amplitude of quantum fluctuations arising from the fermions.
In this work we will truncate the expansion (\ref{ansatz}) to the first order in $\ol\psi\psi$
and therefore consider only the potentials $U_0(\phi), U_1(\phi)$.

\subsection{Evaluation of the trace}

We compute here the trace appearing in eq.(\ref{evolgamma}). 
The second derivatives of $\Gamma$ are, for constant field configurations with 
$\phi=\phi_0$ and $\ol\psi\psi=\omega$,
\bea
\delta^2\Gamma_{\ol\psi\psi}&=&
\left(-\br p+A_1+4A_2\omega+\cdot\cdot\cdot\right)\delta^5(p+q)\nn
\delta^2\Gamma_{\psi\ol\psi}&=&
\left(-\br p-A_1-4A_2\omega+\cdot\cdot\cdot\right)\delta^5(p+q)\nn
\delta^2\Gamma_{\phi\phi}&=&
\left(p^2-A_0^{''}-A_1^{''}\omega+\cdot\cdot\cdot\right)\delta^5(p+q)\nn
\delta^2\Gamma_{\ol\psi\phi}&=&
-\psi\left(A_1^{'}+2A_2^{'}\omega+\cdot\cdot\cdot\right)\delta^5(p+q)\nn
\delta^2\Gamma_{\psi\phi}&=&
\ol\psi\left(A_1^{'}+2A_2^{'}\omega+\cdot\cdot\cdot\right)\delta^5(p+q)\nn
\delta^2\Gamma_{\ol\psi\ol\psi}&=&
\left(-2\psi\psi A_2+\cdot\cdot\cdot\right)\delta^5(p+q)\nn
\delta^2\Gamma_{\psi\psi}&=&
\left(-2\ol\psi\ol\psi A_2+\cdot\cdot\cdot\right)\delta^5(p+q),
\eea
where we denote $A_n=U_n(\phi_0)$ and a prime denotes a derivative with respect to $\phi$. We now write
\be
\partial^2\Gamma=(\Delta+\Xi)\delta^5(p+q),
\ee
where $\Delta$ is the $\phi$-dependent diagonal part of $\partial^2\Gamma$ and $\Xi$ contains
the diagonal and off diagonal terms including the fermions. We have, within the truncation we
consider here (where $A_0$ and $A_1$ only are non vanishing):
\bea
\Delta&=&\begin{pmatrix}
-\br p+A_1 & 0 & 0 \\
0 & -\br p-A_1 & 0 \\
0 & 0 & p^2-A_0^{''}
\end{pmatrix}\nn
\Xi&=&\begin{pmatrix}0 & 0 & -A_1^{'}\psi \\
    0 & 0 & A_1^{'}\ol\psi \\
    A_1^{'}\ol\psi & -A_1^{'}\psi & -A_1^{''}\omega
\end{pmatrix}.
\eea
The inverse of $\partial^2\Gamma$ is obtained by expanding in powers of the fermion field:
\be
\left(\partial^2\Gamma\right)^{-1}=
\delta^5(p+q)\left(\Delta^{-1}-\Delta^{-1}\Xi\Delta^{-1}+
\Delta^{-1}\Xi\Delta^{-1}\Xi\Delta^{-1}+\cdot\cdot\cdot\right).
\ee
We find then
\be
\left(\partial^2\Gamma\right)^{-1}_{\ol\psi\psi}=
\frac{\delta^5(p+q)}{-\br p+A_1}
+\frac{\ol\psi\psi~(A_1^{'})^2~\delta^5(p+q)}{(-\br p+A_1)^2(p^2-A_0^{''})}
+\cdot\cdot\cdot,
\ee
where the dots denote higher orders in the fermion field.
The 5-momentum $p$ is such that $p^2=k^2+l^2/R^2$, where $k$ is the 4-momentum and $l\in{\cal Z}$ labels the
Kaluza Klein modes. The trace in (\ref{evolgamma}) is then (for Euclidean $k$)
\bea\label{trace1}
&&i\mbox{Tr}\left\{(\delta^2\Gamma)^{-1}_{\ol\psi\psi}\right\}\\
&=&L^4\int\frac{d^4k}{(2\pi)^4}\sum_{l\in{\cal Z}}\left\{\frac{-8A_1}{k^2+\frac{l^2}{R^2}+A_1^2}+
\frac{\omega(A_1^{'})^2\left(A_1^2-k^2-\frac{l^2}{R^2}\right)}
{(k^2+\frac{l^2}{R^2}+A_1^2)^2(k^2+\frac{l^2}{R^2}+A_0^{''})}\right\}\nn
&=&L^4\int\frac{d^4k}{(2\pi)^4}\sum_{l\in{\cal Z}}\left\{
\frac{-8A_1}{k^2+\frac{l^2}{R^2}+A_1^2}+\omega(A_1^{'})^2
\frac{\frac{A_0^{''}+A_1^2}{(A_0^{''}-A_1^2)^2}}{k^2+\frac{l^2}{R^2}+A_0^{''}}\right.\nn
&&~~~~~~~~~~~~~~\left.-\omega(A_1^{'})^2\frac{\frac{A_0^{''}+A_1^2}{(A_0^{''}-A_1^2)^2}}{k^2+\frac{l^2}{R^2}+A_1^2}
+\omega(A_1^{'})^2\frac{\frac{2A_1^2}{A_0^{''}-A_1^2}}{(k^2+\frac{l^2}{R^2}+A_1^2)^2}\right\},\nonumber
\eea
where $L^4=\delta^4(0)$ is the volume of the 4-dimensional large space time.
Note the factor 8 in front
of the $\omega$-independent term, corresponding to the trace over Dirac indices of the unity matrix,
in the reducible representation that we consider here. This factor does not appear in the
$\omega$-dependent terms since $\omega$ already takes into account the summation over Dirac indices.
The summation over Kaluza Klein modes is done using the relation \cite{GR}
\be
 \sum_{l\in{\cal Z}}\frac{1}{l^2+a^2}=\frac{\pi}{a}\coth(\pi a),
\ee
as well as its derivative with respect to $a^2$:
\be
\sum_{l\in{\cal Z}}\frac{1}{(l^2+a^2)^2}=\frac{\pi^2}{2a^2}\coth^2(\pi a)+
\frac{\pi}{2a^3}\coth(\pi a)-\frac{\pi^2}{2a^2}.
\ee
The trace (\ref{trace1}) is then
\bea
&&\frac{i}{L^4}\mbox{Tr}\left\{(\delta^2\Gamma)^{-1}_{\ol\psi\psi}\right\}\\
&=&\pi R\left(-8A_1-\omega(A_1^{'})^2\frac{A_0^{''}+A_1^2}{(A_0^{''}-A_1^2)^2}\right)
\int\frac{d^4k}{(2\pi)^4}\frac{\coth(\pi R\sqrt{k^2+A_1^2})}{\sqrt{k^2+A_1^2}}\nn
&&+\pi R\omega(A_1^{'})^2\frac{A_0^{''}+A_1^2}{(A_0^{''}-A_1^2)^2}
\int\frac{d^4k}{(2\pi)^4}\frac{\coth(\pi R\sqrt{k^2+A_0^{''}})}{\sqrt{k^2+A_0^{''}}}\nn
&&+\pi R^2\omega(A_1^{'})^2\frac{A_1^2}{A_0^{''}-A_1^2}
\int\frac{d^4k}{(2\pi)^4}\left(\pi\frac{\coth^2(\pi R\sqrt{k^2+A_1^2})}{k^2+A_1^2}\right.\nn
&&~~~~~~~~~~~~~~~~~~~~~~\left.+\frac{\coth(\pi R\sqrt{k^2+A_1^2})}{R(k^2+A_1^2)^{3/2}}-
\frac{\pi}{k^2+A_1^2}\right),\nonumber
\eea
The evolution equations for the potentials are obtained by identification with the left hand
side of the evolution equation (\ref{evolgamma}), where 
\be
\Gamma=-2\pi RL^4(A_0+A_1\omega),
\ee 
for the constant configuration that we consider. The integrations over $k$ give, for any $A$,
\bea
&&\int\frac{d^4k}{(2\pi)^4}\frac{\coth(\pi R\sqrt{k^2+A^2})}{\sqrt{k^2+A^2}}\\\
&=&\frac{1}{8\pi^5R^3}\int_{\pi RA}^{P} du\left(u^2-(\pi RA)^2\right)\coth(u)\nn
&=&\frac{1}{8\pi^5R^3}\left(I(P)-I(\pi RA)
-(\pi RA)^2\ln\left(\frac{\sinh(P)}{\sinh(\pi RA)}\right)\right)\nonumber
\eea
where
\be\label{deffI}
I(z)=\int_0^z du~u^2\coth(u)~~~~\mbox{and}~~~~P=\pi R\sqrt{\Lambda^2+A^2}.
\ee
Note that $I(z)\simeq z^3/3$ when $z\to\infty$.
We also have
\bea
&&\int\frac{d^4k}{(2\pi)^4}\left(\pi\frac{\coth^2(\pi R\sqrt{k^2+A^2})}{k^2+A_1^2}
+\frac{\coth(\pi R\sqrt{k^2+A^2})}{R(k^2+A^2)^{3/2}}-\frac{\pi}{k^2+A^2}\right)\nn
&=&\frac{1}{8\pi^3 R^2}\int_{\pi RA}^{P} du \left(1-\frac{(\pi RA)^2}{u^2}\right)
\left(u\coth^2(u)+\coth(u)-u\right)\nn
&=&\frac{1}{8\pi^3R^2}\Bigg(2\ln\left(\frac{\sinh(P)}{\sinh(\pi RA)}\right)
-\frac{(\pi R\Lambda)^2}{P}\coth(P)\Bigg)
\eea

\subsection{Evolution equations for the potentials}

Using the previous results, we find for the scalar potential $U_0$:
\be\label{evolU0}
\dot U_0=\frac{m_0U_1}{2\pi^5R^3}\left\{I(P)-I(\pi RU_1)
-(\pi RU_1)^2\ln\left(\frac{\sinh(P)}{\sinh(\pi RU_1)}\right)\right\},
\ee
where $\Lambda$ is an UV cut off, the function $I(z)$ is given in eq.(\ref{deffI}), and
\be
P=\pi R\sqrt{\Lambda^2+U_1^2}.
\ee
We find for the Yukawa potential $U_1$:
\bea\label{evolU1}
&&\dot U_1-m_0\\
&=&\frac{m_0(U_1^{'})^2}{16\pi^5R^3}\frac{U_1^2+U_0^{''}}{(U_1^2-U_0^{''})^2}
\Bigg\{I(P)-I(\pi RU_1)-I(Q)+I\left(\pi R(U_0^{''})^{1/2}\right)\nn
&&-(\pi RU_1)^2\ln\left(\frac{\sinh(P)}{\sinh(\pi RU_1)}\right)
+\pi^2R^2U_0^{''}\ln\left(
\frac{\sinh(Q)}{\sinh\left(\pi R(U_0^{''})^{1/2}\right)}\right)\Bigg\}\nn
&&+\frac{m_0(U_1^{'})^2}{16\pi^3R}\frac{U_1^2}{U_1^2-U_0^{''}}
\Bigg\{2\ln\left(\frac{\sinh(P)}{\sinh(\pi RU_1)}\right)
-\frac{(\pi R\Lambda)^2}{P}\coth(P)\Bigg\}\nonumber
\eea
where 
\be
Q=\pi R\sqrt{\Lambda^2+U_0^{''}}.
\ee
Finally, the evolutions equations (\ref{evolU0}) and (\ref{evolU1}) have to be integrated from 
$\lambda=\infty$, where the fermions do not play a role, to a finite value of $\lambda$, 
where the full interactions between the fermions and the scalar are present.

\section{On the scalar mass}

Since the aim of this article is to describe dynamical mass generation for the fermions,
we will consider a massless scalar field, so as not to have a bosonic contribution to the
fermion mass generated by the interaction with the scalar field. 
Also, this situation can be seen as a toy model for a gauge field coupled to the fermions, even
if no gauge symmetry is present here. 
We show in this subsection, at one-loop, 
that it is indeed possible to have a massless scalar field, provided
a constraint holds on the Yukawa coupling and the scalar self-coupling.

We start with the bare potentials
\bea
U_0^B(\phi)&=&a\phi+\frac{M_0^2}{2}\phi^2+\frac{\eta_0}{24}\phi^4\nn
U_1^B(\phi)&=&\lambda m_0+g_0\phi,
\eea
where the linear term $a\phi$ will cancel the tadpole diagram generated by the 
Yukawa interaction. For these bare potentials, no cubic self-interaction for the scalar field
is generated at one-loop, such that it is enough to compute  
the one-loop scalar mass term, in order to discuss the vacuum of the scalar field. 

The details of the computations are given in the subsection 2.2, and we write here the main steps only.
The one loop correction to the scalar mass term, arising from the boson loop, is
\bea
&&(M^{(1)}_b)^2\\
&=&\frac{i\eta_0}{2\pi R}\int\frac{d^4p}{(2\pi)^4}\sum_n\frac{1}{p^2-(n/R)^2-M_0^2}\nn
&=&-\frac{\eta_0 R}{2\pi}\int\frac{d^4p_E}{(2\pi)^4}\sum_n\frac{1}{n^2+R^2(p_E^2+M_0^2)}\nn
&=&-\eta_0\pi R\int\frac{d^4p_E}{(2\pi)^4}\frac{\coth\left(\pi R\sqrt{p_E^2+M_0^2}\right)}{\sqrt{p_E^2+M_0^2}}\nn
&=&\frac{-\eta_0}{16\pi^5R^3}\left(I(Q_0)-I(\pi RM_0)
-(\pi RM_0)^2\ln\left(\frac{\sinh(Q_0)}{\sinh(\pi RM_0)}\right)\right),\nonumber
\eea
where $p_E$ is the Euclidean momentum and $Q_0=\pi R\sqrt{\Lambda^2+M_0^2}$. 
So as to see the divergences explicitly, one can use the following expansion 
\be
I(Q_0)=I(\pi R\Lambda)+\frac{(\pi RM_0)^2}{2}\pi R\Lambda+{\cal O}(M_0/\Lambda),
\ee
where $I(z)\simeq z^3/3$ when $z\to\infty$. We obtain then
\be
(M^{(1)}_b)^2=\frac{-\eta_0}{16\pi^2}\left(\frac{\Lambda^3}{3}-\frac{M_0^2}{2}\Lambda+~\mbox{finite}\right).
\ee
The one-loop correction to the scalar mass, arising from the fermion loop, is 
\be
(M^{(1)}_f)^2=\frac{(ig_0)^2}{2\pi R}\mbox{tr}\int\frac{d^4p}{(2\pi)^4}\sum_n \frac{-i}{(\br k-\lambda m_0)^2},
\ee
where $k^2=p^2-(n/R)^2$ and the trace will generate a factor 8 coming from the $8\times 8$ gamma matrices.
We have then
\bea\label{Mf2}
&&(M^{(1)}_f)^2\\
&=&\frac{4g_0^2R}{\pi}\int\frac{d^4p_E}{(2\pi)^4}\sum_n\frac{1}{n^2+R^2(p_E^2+(\lambda m_0)^2)}\nn
&&-\frac{8g_0^2R^3(\lambda m_0)^2}{\pi}\int\frac{d^4p_E}{(2\pi)^4}
\sum_n\frac{1}{\left(n^2+R^2(p_E^2+(\lambda m_0)^2\right)^2}\nn
&=&4g_0^2\int\frac{d^4p_E}{(2\pi)^4}
\frac{\coth\left(\pi R\sqrt{p_E^2+(\lambda m_0)^2}\right)}{\sqrt{p_E^2+(\lambda m_0)^2}}\nn
&&-4\pi g_0^2R(\lambda m_0)^2\int\frac{d^4p_E}{(2\pi)^4}
\frac{\coth^2\left(\pi R\sqrt{p_E^2+(\lambda m_0)^2}\right)}{p_E^2+(\lambda m_0)^2}\nn
&&-4g_0^2R(\lambda m_0)^2\int\frac{d^4p_E}{(2\pi)^4}
\left(\frac{\coth\left(\pi R\sqrt{p_E^2+(\lambda m_0)^2}\right)}{R\left(p_E^2+(\lambda m_0)^2\right)^{3/2}}
-\frac{\pi}{p_E^2+(\lambda m_0)^2}\right)\nn
&=&\frac{g_0^2}{2\pi^5R^3}\Bigg(I(P_0)-I(\pi R\lambda m_0)
-3(\pi R\lambda m_0)^2\ln\left(\frac{\sinh(P_0)}{\sinh(\pi R\lambda m_0)}\right)\nn
&&~~~~~~~~~~~+(\pi R\lambda m_0)^2\frac{(\pi R\Lambda)^2}{P_0}\coth(P_0)\Bigg),\nonumber
\eea
where $P_0=\pi R\sqrt{\Lambda^2+(\lambda m_0)^2}$. 
An expansion of $(M^{(1)}_f)^2$ in powers of $\Lambda$ gives
\be
(M^{(1)}_f)^2=\frac{g_0^2}{2\pi^2}\left(\frac{\Lambda^3}{3}-\frac{3}{2}(\lambda m_0)^2\Lambda+~\mbox{finite}\right).
\ee
As expected, the dominant contributions from the boson and fermion loops have opposite signs, and the one-loop scalar mass
is finally
\be
(M^{(1)})^2=M_0^2-\frac{\eta_0}{16\pi^2}\left(\frac{\Lambda^3}{3}-\frac{M_0^2}{2}\Lambda\right)
+\frac{g_0^2}{2\pi^2}\left(\frac{\Lambda^3}{3}-\frac{3}{2}(\lambda m_0)^2\Lambda\right)+~\mbox{finite}.
\ee
So as to cancel $M^{(1)}$, one needs to choose the bare mass term such that
\be
M_0^2\left(1+\frac{\eta_0\Lambda}{32\pi^2}\right)=
\frac{\Lambda^3}{48\pi^2}(\eta_0-8g_0^2)+\frac{3\Lambda}{4\pi^2}g_0^2(\lambda m_0)^2+~\mbox{finite}.
\ee
One can see from this last expression that, in order not to generate a spontaneous symmetry breaking with
the bare mass term, it is necessary that
\be
\eta_0\ge 8g_0^2.
\ee
Finally, the effective scalar potential we will consider in the following of this paper is
\be\label{renormU0}
U_0(\phi)=\frac{\eta}{24}\phi^4+~\mbox{higher powers of}~\phi,
\ee                                                     
where $\eta$ is the effective coupling corresponding to the interaction $\phi^4$ and the higher order
couplings are generated by quantum corrections. With the scalar potential (\ref{renormU0}), the restoration force
for the scalar field is provided by the interaction $\phi^4$, keeping the vacuum at $<\phi>=0$.

As a last remark, 
one can note that the expression (\ref{Mf2}) is consistent with the evolution equation (\ref{evolU0}).
For this, one can observe that the right hand side of eq.(\ref{evolgamma}) is proportional to $\hbar$,
such that replacing $U_0,U_1$ by the
bare potentials $U_0^B,U_1^B$ on the right hand side of eqs.(\ref{evolU0}) and (\ref{evolU1}) leads to the
one-loop $\lambda$-evolution for the potentials.
The one-loop approximation for the $\lambda$-evolution of the scalar potential is then
\be\label{oneloopU0}
\dot U_0^{(1)}=\frac{m_0U_1^B}{2\pi^5R^3}\left\{I(P^B)-I(\pi RU_1^B)
-(\pi RU_1^B)^2\ln\left(\frac{\sinh(P^B)}{\sinh(\pi RU_1^B)}\right)\right\},
\ee
From the latter equation, one can compute the one-loop evolution to the scalar mass term
by making an expansion in $\phi$.
The identification of the power $\phi^2$ on both sides of eq.(\ref{oneloopU0}) gives then
\bea
M^{(1)}\dot M^{(1)}&=&
\frac{\lambda m_0^2g_0^2}{2\pi^3R}\Bigg\{\lambda\pi Rm_0\coth(\lambda\pi Rm_0)
+\frac{3}{2}P_0\coth(P_0)\\
&&~~~~~~~~~~~-3(\lambda\pi Rm_0)^2\frac{\coth(P_0)}{P_0}-3\ln\left(\frac{\sinh(P_0)}{\sinh(\lambda\pi Rm_0)}\right)\nn
&&~~~~~~~~~~~+\frac{(\lambda\pi Rm_0)^4}{2P_0^2}\left(\coth^2(P_0)-1+\frac{\coth(P_0)}{P_0}\right)\nn
&&~~~~~~~~~~~-\frac{(\lambda\pi Rm_0)^2}{2}\big(\coth^2(P_0)-1\big)\Bigg\}.\nonumber
\eea
We checked that this last expression can indeed be obtained by taking the derivative of $(M^{(1)}_f)^2$
(see eq.(\ref{Mf2})) with respect to $\lambda$.

\section{Four dimensional one-loop Yukawa coupling}

We show in this section that the one-loop $\lambda$-evolution of the Yukawa coupling
leads to the well known one-loop coupling, as expected from this method.

As explained in the previous section, 
the one-loop approximation for the evolution of the Yukawa potential is obtained by
replacing $U_0,U_1$ by $U_0^B,U_1^B$ on the right hand side of eq.(\ref{evolU1}), which leads to
\bea\label{oneloopU1}
&&\dot U_1^{(1)}-m_0\\
&=&\frac{m_0(U_1^{B'})^2}{16\pi^5R^3}\frac{(U_1^B)^2+U_0^{B''}}{[(U_1^B)^2-U_0^{B''}]^2}
\Bigg\{I(P^B)-I(\pi RU_1^B)-I(Q^B)\nn
&&~~~~~+I\left(\pi R(U_0^{B''})^{1/2}\right)
-(\pi RU_1^B)^2\ln\left(\frac{\sinh(P^B)}{\sinh(\pi RU_1^B)}\right)\nn
&&~~~~~+\pi^2R^2U_0^{B''}\ln\left(
\frac{\sinh(Q^B)}{\sinh\left(\pi R(U_0^{B''})^{1/2}\right)}\right)\Bigg\}\nn
&&+\frac{m_0(U_1^{B'})^2}{16\pi^3R}\frac{(U_1^B)^2}{(U_1^B)^2-U_0^{B''}}
\Bigg\{2\ln\left(\frac{\sinh(P^B)}{\sinh(\pi RU_1^B)}\right)
-\frac{(\pi R\Lambda)^2}{P^B}\coth(P^B)\Bigg\},\nonumber
\eea
where $P^B=\pi R\sqrt{\Lambda^2+(U_1^B)^2}$ and $Q^B=\pi R\sqrt{\Lambda^2+U_0^{B''}}$.
In this approximation, the Yukawa coupling $g^{(1)}$ can then easily be obtained by noting that
\be
g^{(1)}=\partial_\phi U_1^{(1)}(0)
=\frac{\partial U^{(1)}_1}{\partial U^B_1}\partial_\phi U^B_1(0)
+\frac{\partial U^{(1)}_1}{\partial U^B_0}\partial_\phi U^B_0(0).
\ee
If we take
\bea
U^B_0(\phi)&=&a\phi+\frac{M_0^2}{2}\phi^2+\frac{\eta_0}{24}\phi^4\nn
U^B_1(\phi)&=&\lambda m_0+g_0\phi,
\eea
we have then, at one-loop 
\be
g^{(1)}=\frac{\partial U^{(1)}_1}{\partial U^B_1}\partial_\phi U^B_1(0)
=\frac{g_0}{m_0}\frac{\partial U^{(1)}_1}{\partial U^B_1}\dot U^B_1(0)
=\frac{g_0}{m_0}\dot U_1^{(1)}(0),
\ee
since the linear term $a\phi$ has to be taken into account at the next order ($a$ is already of order 
$\hbar$: it was introduced in order to cancel the tadpole diagram).
The four dimensional Yukawa coupling is defined by
\be\label{defkappa}
\kappa^2=\frac{g^2}{2\pi R},
\ee
and we obtain then, from eq.(\ref{oneloopU1}), the following one-loop coupling
\bea\label{kappa1loop}
&&\kappa^{(1)}-\kappa_0\\
&=&\frac{\kappa_0^3}{8\pi^4R^2}\frac{(\lambda m_0)^2+M_0^2}{[(\lambda m_0)^2-M_0^2]^2}
\Bigg\{I(P_0)-I(\lambda\pi Rm_0)-I(Q_0)+I(\pi RM_0)\nn
&&-(\lambda\pi Rm_0)^2\ln\left(\frac{\sinh(P_0)}{\sinh(\lambda\pi Rm_0)}\right)+
(\pi RM_0)^2\ln\left(\frac{\sinh(Q_0)}{\sinh(\pi RM_0)}\right)\Bigg\}\nn
&+&\frac{\kappa_0^3}{8\pi^2}\frac{(\lambda m_0)^2}{(\lambda m_0)^2-M_0^2}\Bigg\{
2\ln\left(\frac{\sinh(P_0)}{\sinh(\lambda\pi Rm_0)}\right)-\frac{(\pi R\Lambda)^2}{P_0}\coth(P_0)\Bigg\},\nonumber
\eea
where $P_0=\pi R\sqrt{\Lambda^2+(\lambda m_0)^2}$ and $Q_0=\pi R\sqrt{\Lambda^2+M_0^2}$.

The result (\ref{kappa1loop}) can be obtained by a standard one-loop Feynman graph
representing the vertex for zero incoming momentum:
\be
\kappa^{(1)}=\kappa_0+(-i\kappa_0)^3\int\frac{d^4q}{(2\pi)^4}\sum_n\frac{i^2}{(p_n^2-M_0^2)(-\br p_n+\lambda m_0)^2},
\ee
where $p_n^2=q^2+n^2/R^2$. The computation of this graph is a specific case of a more 
general computation done in the subsection 2.2 for the
derivation of the result (\ref{evolU1}), and which leads to eq.(\ref{kappa1loop}).

It is interesting to look at the limit $\Lambda\to\infty$, {\it for fixed control parameter $\lambda$ and
fixed radius $R$}, in order to show explicitly the divergence in the Yukawa coupling.
We use for this the large-$\Lambda$ expansion, obtained with a Taylor expansion of $I(P_0)$ and $I(Q_0)$ around
$I(\pi R\Lambda)$ 
\bea\label{expIlarge}
I(P_0)&=&I(\pi R\Lambda)+\frac{(\lambda\pi Rm_0)^2}{2}\pi R\Lambda+{\cal O}(\lambda m_0/\Lambda)\nn
I(Q_0)&=&I(\pi R\Lambda)+\frac{(\pi RM_0)^2}{2}\pi R\Lambda+{\cal O}(M_0/\Lambda).
\eea
We obtain then from eq.(\ref{kappa1loop})
\bea\label{kappa1loopinfty}
&&\kappa^{(1)}_{\Lambda\to\infty}-\kappa_0\\
&=&\frac{\kappa_0^3}{16\pi}R\Lambda
+\frac{\kappa_0^3}{8\pi^2}\frac{(\lambda m_0)^2+M_0^2}{[(\lambda m_0)^2-M_0^2]^2}
\Bigg\{\frac{I(\pi RM_0)-I(\lambda\pi Rm_0)}{(\pi R)^2}\nn
&&~~~~~~~~~~+(\lambda m_0)^2\frac{3M_0^2-(\lambda m_0)^2}{(\lambda m_0)^2+M_0^2}\ln\left(2\sinh(\lambda\pi Rm_0)\right)\nn
&&~~~~~~~~~~~~~~~~~~~~~~~~~~~-M_0^2\ln\left(2\sinh(\pi RM_0)\right)\Bigg\},\nonumber
\eea
which shows a linear divergence in $\Lambda$. 

Note that, in the precedent expressions, the apparent singularity when $\lambda m_0\to M_0$ is 
canceled by the numerators. As an example, we can check 
eq.(\ref{kappa1loopinfty}), where we note $\lambda m_0=M_0+\epsilon$: an expansion up to $\epsilon^2$ 
gives
\bea
&&\kappa^{(1)}_{\lambda m_0\to M_0}-\kappa_0\\
&=&\frac{\kappa_0^3}{16\pi}R\Lambda-\frac{\kappa_0^3}{16\pi}
\Big\{\pi RM_0\coth(\pi RM_0)+2\ln\left(2\sinh(\pi RM_0)\right)\Big\}+{\cal O}(\epsilon),\nonumber  
\eea
which shows that indeed, no singularity occurs in the limit $\epsilon\to 0$.

\subsection{Limit of the classical system: $\lambda\to\infty$}

At this point, we can check the consistency of the general approach, and show that 
the classical coupling is indeed recovered in the limit $\lambda\to\infty$, {\it for 
fixed cut off $\Lambda$ and fixed radius $R$}.  
We use for this the following large-$\lambda$ expansion, obtained with a Taylor expansion of $I(P_0)$
around $I(\lambda\pi Rm_0)$:
\be\label{expI}
I(P_0)=I(\lambda\pi Rm_0)+\frac{(\pi R\Lambda)^2}{2}\lambda\pi Rm_0+{\cal O}(1/\lambda).
\ee
It can easily be seen from eq.(\ref{kappa1loop}) that we obtain then
\be
\lim_{\lambda\to\infty}\kappa^{(1)}=\kappa_0.
\ee
As expected, an infinite massive system does not generate quantum fluctuations and 
remains classical. As $\lambda$ decreases, quantum fluctuations gradually appear in the system
and the scale $\lambda m_0$ plays the role of the typical energy scale that characterizes the 
system.

\subsection{Four dimensional limit: $R\to 0$}

The limit $R\to 0$, {\it for fixed cut off $\Lambda$ and fixed control parameter $\lambda$}, 
is found from eq.(\ref{kappa1loop}) if we use the small-$z$ expansion obtained with a Taylor
expansion of $u^2\coth(u)$ around $u=0$:
\be\label{expIsmall}
I(z)=\frac{z^2}{2}+\frac{z^4}{12}+{\cal O}(z)^6,
\ee
which gives,
\bea\label{kappa1loopR0}
\kappa^{(1)}_{R\to 0}
&=&\kappa_0+\frac{\kappa_0^3}{16\pi^2}\Bigg\{
(\lambda m_0)^2\frac{(\lambda m_0)^2-3M_0^2}{[(\lambda m_0)^2-M_0^2]^2}
\ln\left(1+\frac{\Lambda^2}{(\lambda m_0)^2}\right)\nn
&&~~~~~~~~~~~~~+M_0^2\frac{(\lambda m_0)^2+M_0^2}{[(\lambda m_0)^2-M_0^2]^2}\ln\left(1+\frac{\Lambda^2}{M_0^2}\right)\nn
&&~~~~~~~~~~~~~-\frac{2(\lambda m_0)^2\Lambda^2}{[(\lambda m_0)^2+\Lambda^2][(\lambda m_0)^2-M_0^2]}\Bigg\}.
\eea
Therefore, the linear divergence for $R\ne 0$ becomes logarithmic in the limit $R\to 0$, 
which is consistent with the results found in \cite{dienes}.

The expression (\ref{kappa1loopR0}) can be obtained from a standard one-loop computation
using Feynman rules. Indeed, the one-loop correction to the vertex corresponding to a
bare Yukawa coupling $\kappa_0$ and a scalar field of mass $M_0$ is, for vanishing incoming momenta,
\bea
&&(-i\kappa_0)^3\int\frac{d^4q}{(2\pi)^4}\frac{i^2}{(q^2-M_0^2)(-\br q+\lambda m_0)^2}\\
&=&\frac{\kappa^3_0}{16\pi^2}\int d(q_E^2)\frac{q_E^2[q_E^2-(\lambda m_0)^2]}{(q_E^2+M_0^2)[q_E^2+(\lambda m_0)^2]^2}\nn
&=&\frac{\kappa_0^3}{16\pi^2}\Bigg\{M_0^2\frac{(\lambda m_0)^2+M_0^2}{[(\lambda m_0)^2-M_0^2]^2}
\ln\left(1+\frac{\Lambda^2}{M_0^2}\right)\nn
&&~~~~~~~~~+(\lambda m_0)^2\frac{(\lambda m_0)^2-3M_0^2}{[(\lambda m_0)^2-M_0^2]^2}
\ln\left(1+\frac{\Lambda^2}{(\lambda m_0)^2}\right)\nn
&&~~~~~~~~-\frac{2(\lambda m_0)^2\Lambda^2}{[(\lambda m_0)^2-M_0^2][\Lambda^2+(\lambda m_0)^2]}\Bigg\},\nonumber
\eea
where $q_E$ is the Euclidean momentum, which corresponds to the result (\ref{kappa1loopR0}).

\section{Dynamical fermion mass generation}

The presence of a dimensionful parameter, the radius $R$, enables the dynamical
mass generation in this system. We now look at the generation of such a
mass for the fermions, and come back to eq.(\ref{evolU1}), where the fermion mass
evolution is given by $\dot m=\dot U^{(1)}(0)$. Dynamical mass generation can be studied in the
functional technique we describe here by taking the limit $\lambda\to 0$, for finite $m_0$. This was
already done in \cite{moi} for the dynamical mass generation in $QED_3$, where a non-perturbative
relation was found between the dynamical mass and the dressed coupling. 

We stress here that the following derivation goes beyond one-loop, which is possible as a consequence of the 
non-perturbative properties of the evolution equation (\ref{evolgamma}).

\subsection{Derivation of the dynamical mass}

We give here the detailed approximations involved in the derivation of the dynamical mass $m_{dyn}$.
We take the effective scalar potential (\ref{renormU0}) such that $U_0^{''}(0)=0$, 
and we note $U_1^{'}(0)=g$. 
In what follows, we will take the limit $\lambda\to 0$ for fixed $\Lambda$. Keeping the 
dominant terms in $\Lambda$ and using the expansion (\ref{expIlarge}), 
we obtain from eq.(\ref{evolU1}) and in terms of $\kappa$:
\be
\dot m=m_0+\frac{m_0\kappa^2}{8\pi^2}\left\{\frac{\pi R\Lambda}{2}-\frac{I(\pi Rm)}{(\pi Rm)^2}
-\ln[2\sinh(\pi Rm)]\right\}+{\cal O}(m/\Lambda).
\ee
The next step is to suppose that $\pi Rm<<1$, such that an expansion in $\pi Rm$ gives, taking into
account the expansion (\ref{expIsmall}),
\be\label{dotm1}
\dot m=C-\frac{m_0\kappa^2}{8\pi^2}\ln(2\pi Rm)+{\cal O}(\pi Rm)^2,
\ee
where the constant $C$ is
\be
C=m_0\left(1+\frac{\kappa^2R\Lambda}{16\pi}-\frac{\kappa^2}{16\pi^2}\right).
\ee
In order to integrate eq.(\ref{dotm1}), we note that $\dot m=m_0+{\cal O}(\kappa^2)$, such that 
we write eq.(\ref{dotm1}) as
\be
\dot m\simeq C-\frac{\dot m\kappa^2}{8\pi^2}\ln(2\pi Rm),
\ee
and the integration over $\lambda$ gives
\be\label{dotm2}
m\simeq \lambda C-\frac{m\kappa^2}{8\pi^2}\big(\ln(2\pi Rm)-1\big).
\ee
In this integration, we took a vanishing constant of integration since
there should not be any $\lambda$-independent mass.  
We now consider the limit $\lambda\to 0$, where $m=m_{dyn}$. Neglecting 
1 compared to the logarithm in the right hand side of eq.(\ref{dotm2}), 
we finally obtain for the dynamical mass
\be\label{mdyn}
m_{dyn}\simeq\frac{\exp\left(-8\pi^2/\kappa^2\right)}{2\pi R}.
\ee
As expected from a non perturbative process, $m_{dyn}$ cannot be expanded in powers of $\kappa$.

A numerical study of the dynamical mass was done in \cite{abe}, where the authors consider bulk fermions
coupled to brane fermions via a four-fermion interaction. There, a critical
coupling was found for the dynamical mass generation. Other studies involving Abelian gauge
models and a Schwinger-Dyson approach \cite{SDgauge} led to such a non-perturbative dynamical mass, 
small compared to $1/R$, and showed the presence of a critical coupling. Finally, the Randall-Sundrum 
scenario was also studied in connection to dynamical mass generation in \cite{randallsundrum} 
and led to similar conclusions.
                                                                                                                      
In order to conclude about the existence of a critical coupling for the dynamical mass generation,
we should consider a gradient expansion more general than eq.(\ref{ansatz}), taking
into account the momentum dependence of the fermion self energy. This is left for
a future work, but the main point of the present derivation was to obtain the result (\ref{mdyn})
independently of any renormalization scheme: the cut off $\Lambda$ naturally disappears
from the equations in the limit where the control parameter vanishes $\lambda\to 0$,
which is an advantage of the method proposed here.

\subsection{Physical implication}

The dynamical mass (\ref{mdyn}) appears in the four dimensional effective theory,
as the dressed mass of the would-be massless fermionic mode of the Kaluza-Klein tower. To see this, 
we write the fields as a Fourier expansion in the compactified direction:
\bea
\phi(x^\rho,x^4)&=&\frac{1}{\sqrt{2\pi R}}\sum_n\xi_n(x^\rho)\exp\left(i\frac{n}{R}x^4\right)\nn
\psi(x^\rho,x^4)&=&\frac{1}{\sqrt{2\pi R}}\sum_n\chi_n(x^\rho)\exp\left(i\frac{n}{R}x^4\right),
\eea
where the factors $1/\sqrt{2\pi R}$ give the expected mass dimension
for the four dimensional fields, and $\rho=0,...,3$. In the limit $\lambda\to 0$, we have
$U_1\simeq m_{dyn}+g\phi$, the effective action (\ref{ansatz}) reads then 
\bea\label{gamman}
\Gamma_{\lambda=0}&=&\int d^4x \sum_n\left\{\frac{1}{2}\partial_\rho\xi_n\partial^\rho\xi_{-n}
-\frac{n^2}{R^2}\xi_n\xi_{-n}\right.\nn
&&~~~~~~~~~~\left.+i\ol\chi_n\Gamma^\rho\partial_\rho\chi_{-n}
-\frac{n}{R}\ol\chi_n\Gamma^4\chi_{-n}\right\}\nn
&&-\int d^4x\sum_{n,m}\ol\chi_n\chi_m\left(m_{dyn}\delta_{n,m}
+\frac{g}{\sqrt{2\pi R}}\xi_{-n-m}\right)\nn
&&-2\pi R\int d^4x~U_0\left(\frac{1}{\sqrt{2\pi R}}\sum_n\xi_n(x^\rho)e^{inx^4/R}\right),
\eea
where $\rho=0,...,3$. 
The eight-component fermions $\chi_n$ we consider are made of the two four-component
flavours $\chi=(\chi^{(1)},\chi^{(2)})$ (we omit the Kaluza-Klein indice $n$),
where each flavour can be decomposed in terms of the two-component chiralities
$\chi^{(j)}=(\chi^{(j)}_R,\chi^{(j)}_L)$.
$\ol\chi\Gamma^4\chi$ gives then, in 4 dimensions, the following mass terms \cite{gamma},
which are identical for both flavours,
\bea\label{dimred}
\ol\chi\Gamma^4\chi&=&(\chi^{(1)\dagger}~\chi^{(2)\dagger})
\begin{pmatrix}
\gamma^0 & 0 \\ 0 & -\gamma^0 
\end{pmatrix}
\begin{pmatrix}
\gamma^4 & 0 \\ 0 & -\gamma^4
\end{pmatrix}
\begin{pmatrix}
\chi^{(1)} \\ \chi^{(2)}
\end{pmatrix}\nn
&=&\ol\chi^{(1)}\gamma^4\chi^{(1)}+\ol\chi^{(2)}\gamma^4\chi^{(2)}\nn
&=&\chi^{(1)\dagger}_L\chi^{(1)}_R-\chi^{(1)\dagger}_R\chi^{(1)}_L
+\chi^{(2)\dagger}_L\chi^{(2)}_R-\chi^{(2)\dagger}_R\chi^{(2)}_L.
\eea
Note that these mass terms, induced by the dimensional reduction, are pseudo scalars under
a Lorentz transformation, whereas 
the Dirac mass terms, arising from dynamical mass generation, are scalars. Indeed,
a Dirac mass term $\ol\chi\chi$ is 
\bea\label{diracmass}
\ol\chi\chi&=&(\chi^{(1)\dagger}~\chi^{(2)\dagger})
\begin{pmatrix}
\gamma^0 & 0 \\ 0 & -\gamma^0
\end{pmatrix}
\begin{pmatrix}
\chi^{(1)} \\ \chi^{(2)}
\end{pmatrix}\nn
&=&\ol\chi^{(1)}\chi^{(1)}-\ol\chi^{(2)}\chi^{(2)}\nn
&=&\chi^{(1)\dagger}_L\chi^{(1)}_R+\chi^{(1)\dagger}_R\chi^{(1)}_L
-\chi^{(2)\dagger}_L\chi^{(2)}_R-\chi^{(2)\dagger}_R\chi^{(2)}_L,
\eea
and has a different sign for each flavour.

If we neglect quantum effects arising from the massive modes,
with masses $n/R$, the theory obtained is equivalent to setting $\xi_n=\chi_n=0$ for $n\ne 0$ in 
eq.(\ref{gamman}). 
The four dimensional effective action for the massless mode $\xi_0$ and the would-be massless mode $\chi_0$ is therefore
\bea
\Gamma_{\lambda=0}^{(0)}
&=&\int d^4x\Big\{\frac{1}{2}\partial_\mu\xi_0\partial^\mu\xi_0+i\ol\chi_0\br\partial\chi_0\nn
&&~~~~~~~~~~~~-m_{dyn}\ol\chi_0\chi_0-\kappa\xi_0\ol\chi_0\chi_0-V(\xi_0)\Big\},
\eea
where 
\be
V(\xi_0)=2\pi R\times U_0\left(\frac{\xi_0}{\sqrt{2\pi R}}\right).
\ee
The latter potential is independent of $R$ at the tree level, what can be seen after a redefinition of the coupling 
constants present in $U_0$ (these redefinitions are similar to the one 
given in eq.(\ref{defkappa})).

Thus it is possible to observe a small mass, compare to $1/R$, in the effective four
dimensional spectrum, since $Rm_{dyn}<<1$ for a small Yukawa coupling.

\section{Conclusion and outlook}

Using an exact functional method that was already successfully applied to other models 
(\cite{scalar,QED,moi}), we studied in the Kaluza Klein context the evolution of a toy Yukawa model with the 
amplitude of quantum fluctuations, controlled by the bare fermion mass. We showed the consistency
of the approach by comparing our results with well-known one-loop results. Beyond one loop,
and using our non-perturbative evolution equations, we were able to compute the fermion mass 
dynamically generated in this system and could derive a cut off-independent expression.

We plan to study the full evolution equations (\ref{evolU0}) and (\ref{evolU1}) by assuming an
expansion of the potentials in powers of the scalar field. The system of differential equations that
are then obtained can only be treated numerically, which is the topic of a new article. We hope then
to study in details the non-perturbative implications of the compactified dimension on the effective 
four dimensional theory, and more specifically on the dynamical mass generation. Also, by considering a
more advanced gradient expansion for the effective action, it is possible to study the occurrence of
a critical coupling for the dynamical mass generation, and hence study the corresponding phase transition. 
Finally, a supersymmetric generalization of the present model is possible by considering the 
same bare mass $\lambda m_0$ for the scalar as well as for the fermion. The evolution of the
effective action will then contain an additional contribution, that will cancel 
quantum corrections to the potential terms, including the Yukawa interaction, due to non-renormalization
theorems. In this situation, it would be interesting to study wave function renormalization effects,
and also susy breaking by allowing different functions of $\lambda$ for the bare masses. 

\vspace{0.5cm}

\nin{\bf Acknowledgments:} This work is supported by the EPEAEK program "Pythagoras", and co-funded by
the European Union (75\%) and the Hellenic State (25\%).

\end{document}